# Flexible IoT Datapath Programming using P4

Shahzad
*Electronics and Computer Engineering Department*
Hongik University
Sejong, South Korea
shazadrocks@gmail.com

Eun-Sung Jung
*Department of Software and Communications Engineering*
Hongik University
Sejong, South Korea
ejung@hongik.ac.kr

*Abstract*—The progress of the network and device technologies enables any device to be connected to the Internet thereby forming an Internet of Things (IoT) architecture. Internet-related activities based on the IoT services can inevitably generate a lot of real-time data which leads a big data phenomenon. To efficiently handle such big IoT data, software-defined networking (SDN) techniques could come into play. Among SDN techniques, P4 is a high-level language for programming protocol-independent packet processors. This paper presents our preliminary work on how we can use P4 to achieve flexible datapath programming for large scale IoT streaming data streaming. As a case study, we describe the implementation of dynamic MST-based datapath programing for IoT data collection in an arbitrary network topology.

## I. INTRODUCTION

With rapid advances of network, computing, and device technologies, the Internet has been everywhere which affects significantly on people's living. Especially, not only computers but also any devices also connected to the Internet anytime and anywhere. Internet-related activities based on the Internet-of-Things (IoT) services inevitably generate a lot of real-time data, which leads a big data phenomenon. To efficiently handle such a big IoT data, software-defined networking (SDN) techniques could come into play.

SDN is an intelligent network framework by separating the control plane and the data plane of forwarding devices [1]. As you can see in Fig. 1, with the separation of the control and data plane, we need interfaces between controller platform and other components, i.e. northbound applications and southbound devices [2].

P4 is a high level language for programming protocol independent packet processors [3]. Hence the name P4. In short, P4 is an extension of OpenFlow for flexible handling of various protocols. In OpenFlow, packets can be matched only on a predefined set of header fields (e.g. Ethernet, TCP/IPv4, MPLS, etc.). The current version of OpenFlow supports up to 41 different header fields. P4 mainly targets this deficiency of OpenFlow and allows operators to define their own header types. Fig. 2 shows that P4 provides parser and table configuration functionalities in addition to classic OpenFlow for flexible handling of various protocols.

In recent times researchers have shifted from the conventional approaches and switched to P4 for programming IoT devices because of the flexibility and openness that P4 provides [4], [5].

The Open Networking Foundation (ONF) community has a dedicated team called "P4 Brigade" for working on

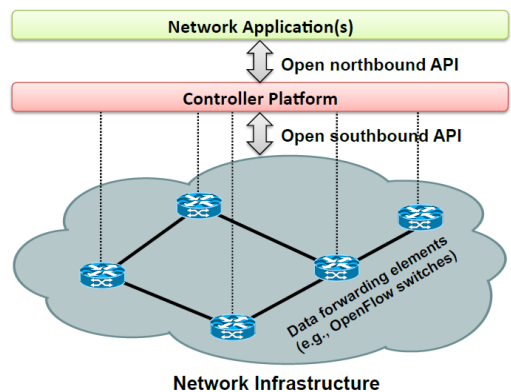

Figure 1. A general SDN architecture

P4 projects [6]. In 2017, a demo, which was a joint effort by google, barefoot networks and ONF, showed ONOS controller and P4Runtime together implementing a load balancing scheme on top of tofino chip based switches. The demo consisted of a datacenter type leaf-spine topology made of four switches that contained the P4 programmable tofino chip from barefoot networks. It is easier in P4 to achieve this as compared to OpenFlow because in OpenFlow variations in data plane pipelines are hard to abstract. On the other hand, P4 enables custom pipelines to be defined.

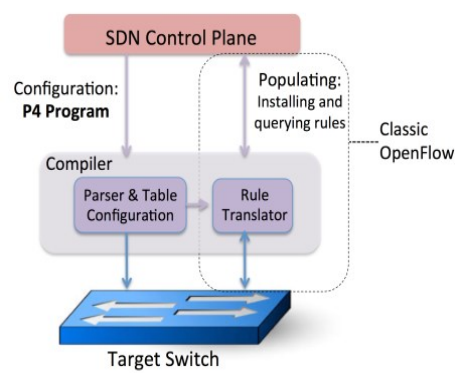

Figure 2. P4 is a language to configure switches from [2].

Authors in [4] presented a new approach to utilize application layer information for QoS improvement in Software Defined Wide Area Networks (SD-WANs). P4 switches are used to intelligently route packets on multipath. P4 switch extracts the Stream ID value from application header and use it to forward the packets on appropriate route.

Our contributions are: (1) we have described what functionalities in the context of SDN would help improve IoT streaming data collection. (2) We have shown that the flexibility of P4 programming could help dynamically establish IoT datapath on users' demand and (3) have also described detailed steps of minimum spanning tree (MST)-based datapath establishment as a case study.

This paper is organized as follows. In Section II, IoT datapath programming using P4 is described, in Section III, a case study of MST is presented, and we conclude in Section IV.

## II. DATA PATH PROGRAMMING USING P4

In this section, we describe how we can define the data path of IoT networks using P4 language.

### A. IoT Streaming data collection requirements

In general, IoT data flows traverse sensors, base stations, Internet, and finally data centers or servers for central processing. In addition, each user has different IoT data requirements such as geographical coverage and data types and associated operations. IoT Streaming data collection requirements can be enumerated as follows.

- Coverage (e.g., area)
- Data type (e.g., temperature)
- Rate (e.g., 10kbps)
- Jitter (e.g., at most 10 ms jitter across multiple IoT devices)
- Operations (e.g., sum or average)

We thus envision that a user can send an IoT streaming request to our IoT datapath programming framework, and in response, our framework establishes the datapath meeting the requirements. For example, a user send the request of the average temperature of Seoul area at the interval of 10 seconds with jitters of 100ms across base stations, the framework may build a MST covering base stations in Seoul area and program average operation on the data collection path to data centers/servers. Authors in [7] present a detailed overview of the routing algorithms used in IoT systems.

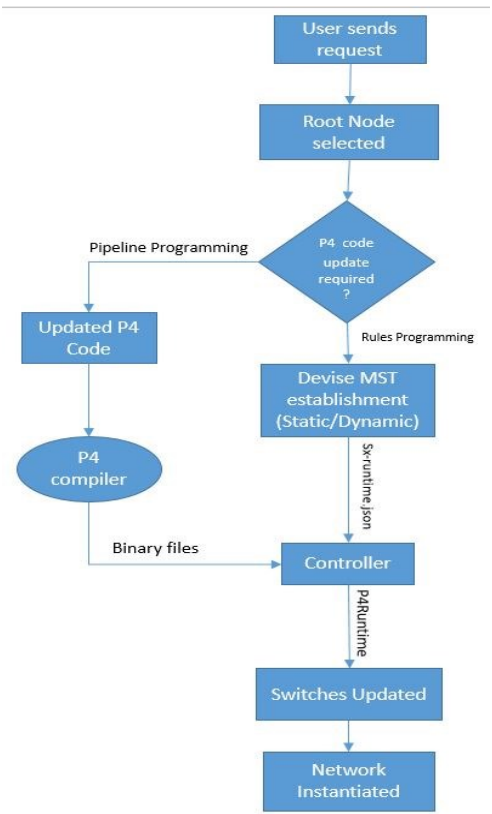

Figure 3. Overall procedures

### B. Workflow for Flexible Data Path Programming

IoT streaming data path programming should meet the above requirements and also utilize the network resources efficiently. In that regard, a MST-based routing is showcased for flexible data path routing in this paper. Given a root node, which is a base station for data collection and forwards collected data to data centers/servers per user demand, we program the switches in the network using P4 language to forward packets to the root node following an MST. The workflow in Fig. 3 is explained below.

1) User sends a request.
2) Given the network topology, one node is selected as the root node for MST.
3) Based on the user input, it is checked whether the switch pipeline programming (P4 code) requires any changes. If any change is required, P4 program is updated accordingly otherwise regular P4 program is used.
4) Now that user inputs are acquired and we have a root node selected, MST is computed in the network topology. This requires new flow rules to be installed in the switches establish MST. This can be done statically (making changes to the flow rule files directly) or dynamically (writing a control plane application). This depends on whether the root node is fixed or changing.
5) The flow rules and P4 program's binary (regular/updated) are forwarded to the central controller.
6) The switches are updated by the controller via P4Runtime API (explained in later section).
7) Finally the Network is instantiated and user request is processed.

### C. P4 Programming

```
/*************************************************************
************* I N G R E S S   P R O C E S S I N G  *********
*************************************************************/

control MyIngress(inout headers hdr,
                  inout metadata meta,
                  inout standard_metadata_t standard_metadata) {
    action drop() {
        mark_to_drop();
    }

    action ipv4_forward(macAddr_t dstAddr, egressSpec_t port) {
        standard_metadata.egress_spec = port;
        hdr.ethernet.srcAddr = hdr.ethernet.dstAddr;
        hdr.ethernet.dstAddr = dstAddr;
        hdr.ipv4.ttl = hdr.ipv4.ttl - 1;
    }

    table ipv4_lpm {
        key = {
            hdr.ipv4.dstAddr: lpm;
        }
        actions = {
            ipv4_forward;
            drop;
            NoAction;
        }
```

Figure 4. Ingress Configuration

Programming with P4 is slightly different from conventional SDN programming. P4 programming can be divided into two parts: Switch pipeline programming and switch flow rules programming.

*1) Switch Pipeline Programming*

Switch pipeline programming refers to the P4 program that is installed into a forwarding device. A P4 program tells a switch how to operate thus defining the switch's pipeline. Some major configurations of the pipeline include header definition, ingress configuration, egress configuration etc. We now explain the ingress configuration in detail.

Ingress configuration decides what will happen to a packet when it arrives at one of the switch's ports. Programming ingress configuration is to dissect the packet received at the switch and do some computation or decision making according to the operator's use case. Fig. 4 shows the ingress configuration code for S1 (switch 1), a snippet from the P4 program installed into S1. In our case, incoming packets are checked for the ipv4 header and longest prefix match is performed from the flow table. Once the packet is identified to be an IP packet, the "dstAddr" field in the packet (i.e. the root node IP address) is matched with the switch flow table to find the appropriate egress port for the packet. As a result the packet is forwarded to the root node following the root node's MST.

*2) Switch Flow Rules Programming*

In addition to switch pipeline programming, we also

```json
{
  "table": "MyIngress.ipv4_lpm",
  "match": {
    "hdr.ipv4.dstAddr": ["10.0.1.1", 32]
  },
  "action_name": "MyIngress.ipv4_forward",
  "action_params": {
    "dstAddr": "00:00:00:00:01:01",
    "port": 1
  }
},
{
  "table": "MyIngress.ipv4_lpm",
  "match": {
    "hdr.ipv4.dstAddr": ["10.0.2.2", 32]
  },
  "action_name": "MyIngress.ipv4_forward",
  "action_params": {
    "dstAddr": "00:00:00:05:05:02",
    "port": 3
  }
},
{
  "table": "MyIngress.ipv4_lpm",
  "match": {
    "hdr.ipv4.dstAddr": ["10.0.3.3", 32]
  },
  "action_name": "MyIngress.ipv4_forward",
  "action_params": {
    "dstAddr": "00:00:00:05:05:02",
    "port": 3
  }
},
```

Figure 5. s1-runtime.json snippet

have to write the flow rules for the switch. Because a p4 program defines a switch's pipeline but the rules within each switch flow table are generated separately and inserted by the control plane.

In order to install the program and rules into a switch, there needs to be an interface between controller and the switch. Using this interface one can install the P4 binary that is generated from the P4 program, flow rules, which in our case are the "sx-runtime.json" files, and other control messages. Currently available interfaces include P4 compiler auto-generated runtime APIs, switch abstraction interface (SAI), P4Runtime API.

D. *Flow Table Installation*

Among several options, we use P4Runtime to program and install flow rules for switches. Depending on dynamics of MST topology, the following two approaches can be taken.

*1) Static flow rule installation*

In Static Flow rules installation, the forwarding rules for each switch are written and maintained in "sx-runtime.json" files where x stands for the switch number. This approach is proactive in nature and can be used to implement static polies. Any changes in the P4 program that add or rename tables, keys, or actions will need to be reflected in these "sx-runtime.json" files. Every switch is configured with a separate "sx-runtime.json" file.

Fig. 5 is a snippet of flow rules for S1. This table corresponds to the IPv4 forwarding in our implementation. A longest prefix match is performed on the destination IP address in each packet and based on that destination IP address, destination mac address and appropriate egress port is selected on the switch for the packet to depart. Since this is the static rules installation, we assume that the root node is fixed and every node in the network knows about the root node. MST establishment is achieved by configuring the flow rules for each switch in such a way that packets follow MST of the root node. It can be seen in Fig. 5 that there is an egress port selection in each switch file for all the nodes in the network. So when a packet, destined for the root node, arrives at a switch it is sent over the egress port which falls in the root node's MST path. In this way the packets traverse through only those switches that are constituents of the root node's MST.

*2) Dynamic flow rules installation*

Although a controller is required in the static flow rule installation also but that is only to push the forwarding rules. So it is considered to be a static controller. In the dynamic flow rule installation, a controller is used to dynamically install the forwarding rules at runtime using the P4Runtime API. E.g. If the Root node is changed in the network, the controller will update the switch rules and re-establish MST at runtime.

III. CONCLUSIONS AND FUTURE WORK

We propose to dynamically program IoT streaming data on demand of users' IoT data collection requirements using P4. Our future work is to completely automate the IoT datapath programming such that multiple users submit IoT data collection requests and in accordance with the requests, our framework establishes IoT datapath from proper base stations to data centers or servers while satisfying users' requirements.

IV. ACKNOWLLEGMENT

This work was supported by Basic Science Research Program through the Ministry of Education of the Republic of Korea and the National Research Foundation of Korea (NRF-2017R1D1A1B03033632).